# Distributed Smart Measurement Architecture for Industrial Automation

Cyber Physical Systems for Europe (CPS4EU)

Paolo Azzoni
*Eurotech Group*
Amaro, Italy
paolo.azzoni@eurotech.com

Gianfranco Caminale
*Leonardo spa*
Genoa, Italy
gianfranco.caminale@leonardocompany.com

Marco Carratù,
University of Salerno
Salerno, Italy
mcarratu@unisa.it

Salvatore Dello Iacono
University of Salerno
Salerno, Italy
sdelloiacono@unisa.it

Giuseppe Fenza
*University of Salerno*
Salerno, Italy
gfenza@unisa.it

Nicola Gallo
*Leonardo spa*
Grottaglie, Italy
nicola.gallo@leonardocompany.com

Consolatina Liguori
*University of Salerno*
Salerno, Italy
tliguori@unisa.it

Elisa Londero
*Eurotech Group*
Amaro, Italy
elisa.londero@eurotech.com

Antonio Pietrosanto
*University of Salerno*
Salerno, Italy
apietrosanto@unisa.it

Nicolo' Rebella
*Leonardo spa*
Genoa, Italy
nicolo.rebella@leonardocompany.com

*Abstract*— Cyber-Physical Systems (CPSs) employed for Industrial Automation often require the adoption of a hybrid data processing approach mediating between cloud, edge, and fog computing paradigms. Nowadays, it is possible to shift data pre-processing capabilities closer to data sensing to collect environmental measurements locally on the edge or deep edge. In line with the emerging computing paradigms, this work proposes a solution that includes both software and hardware components and which simplifies the deployment of smart measurement systems. The solution stresses also the adoption of standards and open data paradigms for simplifying the integration and ensuring the interoperability of all the systems involved. The distributed smart measurement solution has been adopted in an Industry Automation use case included in the project Cyber-Physical Systems for Europe (CPS4EU). The use case attains with monitoring of an industrial trimming machine operating in the production process of a big part of a civil aircraft, where the sensing and processing capabilities of the distributed smart measurement system allow to collect different parameters of work parts to satisfy the expected quality of the production process.

## I. Introduction

The massive amount of data generated by many industrial IoT solutions required to empower monitoring and production processes, stresses traditional computing paradigms. Real-time data processing, analytics and knowledge generation, preserving low latency and high throughput, represent critical constraint to meet. Although a general reference architecture to address big data information systems (Lambda [1] or Kappa Architecture) already exists, its deployment should be broken up into different levels, namely cloud, edge, and fog computing, to satisfy the computational requirements of the Industry Automation.

The analysis of the Industrial Automation use cases involved in the project "Cyber Physical Systems for Europe" (CPS4EU) reveals heterogeneous requirements in terms of data processing. Parameters measurement require IoT-inspired solutions that efficiently exploit the hardware resources and capabilities available on the edge, where the data are generated. At the same time, the huge amount of collected data could enable the extraction of valuable predictive Machine Learning (ML) models requiring high computational power. Thus, hybrid solutions finding the right trade-off between cloud and edge computing are sought for. A layered-based solution distributes data processing among the layers and balances the computing load, increases the efficiency and the autonomy of the computing nodes, and reduces the latency and the bandwidth.

This paper proposes a general architecture (named Distributed Smart Measurement Architecture) and a solution implementing it, which includes software and hardware components for edge computing, actuators and a distributed sensing infrastructure, composed of nodes that acquire real-time parameters from the field (i.e. vibrations, temperature, and other physical quantities). The sensing nodes should be easily integrated, by producing actionable data directly from the field where they are deployed. They adopt MQTT for the connectivity and JSON as a data exchange format. The solution is comliant to ISO/IEC/IEEE 21451.001 standard, for signal pre-processing algorithms running on the edge. Finally, the solution is adopted in a real industrial use case, developed in CPS4EU project and intended to automate the quality monitoring of the trimming process in the production line of a civil aircraft. Physical parameters are collected from the working environment, aggregated on a multi-service gateway (an industry grade edge computing platform named Edge Computing Gateway) and delivered to the cloud for the identification of a model. The model is finally deployed on the multi-service gateway to dynamically suggest the best paramenters for the trimming process.

The paper is organized as follows. Section II introduces the application domain of Industrial CPSs and the main idea behind the project CPS4EU. Section III summarizes the enabling technologies and the related works. Section IV and Section V define, respectively, the proposed Distributed

Smart Measurement Architecture and the role of the Edge Computing Gateway. Section VI illustrates the case study, and the deployment and usage of the smart measurement solution. Conclusions and future works close the paper.

## II. CPS4EU AND INDUSTRIAL CPSS

Cyber Physical Systems represent key enablers to unleash the innovation potential of European industries and their importance is increasing with massive digitalization, opening new market opportunities. Industrial CPSs are the new generation of systems combining sensing, embedded computing and local intelligence, and high-speed connectivity to create a link between the physical and the digital world and to enable cooperation among the systems. The project CPS4EU focuses on this domain and aims at handling its scientific and technical challenges by integrating common software and hardware technological components for CPSs to build efficient pre-integrate architectures that can be used across different products and across several applications. In the project, different application domains like Industry Automation, Automotive, and other industrial sectors are addressed. Coherently with the project ambition, the proposed architecture, adopted in the use case described in Section VI, will likely be re-used to address similar requirements in other use cases and application domains.

## III. RELATED WORKS

During the last decades, industry automation has mostly focused industry automation at any level, from individual machines to entire assembly lines, which are used to chain various processing steps and allow for fully automated production of goods. The goal is to increase the efficiency of the production process, the quality of the products and the overall competitiveness of the industry. The project CPS4EU starts from these motivations and increases the automation level providing pre-integrated components for industry digitalisation. This work aims at designing a Cyber-Physical System [2] that requires a complex and distributed perception layer which relies on distributed smart sensor nodes.

First of all, there is a need to decouple the location of data sensing nodes [3] as suggested by the Reference Architectural Model Industry 4.0 (RAMI 4.0) [4] which builds upon ubiquitous decentralized computing paradigms [5,6]. The decoupling requires that sensing nodes are equipped with micro-controllers or microprocessors (i.e., smart sensors) to increase the processing power at the sensing level [7] and to share the environment perception sending processed information rather than raw data [8]. In this way, the perception layer of CPSs could be realized by deploying sensor networks, that are distributed systems integrating sensor information at the sensing level [9].

Moreover, monitoring industrial processes requires an efficient information sharing of the measures collected by the smat sensors: a good solution for information sharing and integration rely on the publish–subscribe pattern [13] that enables the decoupling of the sensor devices (producing the information) from the other components included in the CPS. According to this approach, [14] proposes a generic solution based on small autonomous wireless sensor nodes, small wireless receivers connected to the Internet, and a cloud architecture that provides data storage and delivery to remote clients. The MoSeS-Pro [15] sensor system combines various micro-sensors with electronics for data acquisition and pre-processing, and communication interfaces.

In addition, solutions adopting distributed sensor systems are applied in multiple contexts. A smart home data management in which a smart gateway collects data generated by home sensors is proposed by Ahsan et al. [16]. Many other examples of distributed systems regard smart grids [17], [18]. Details about an architecture based on distributed sensing systems for smart ports are described in [19].

## IV. DISTRIBUTED SMART MEASUREMENT ARCHITECTURE

A smart solution to manage the computing power required by the sensor networks, composed of a variable number of connected nodes, consists in balancing the computational load over the edge and the cloud/data center. With this approach, depending on the application, it is possible to efficiently distribute the data processing partially on the deep edge (namely in the sensors units where data are pre-processed just after the acquisition), on the edge (in a multiservice gateway deployed close to the sensing infrastructure) and, finally, on the cloud or in a data center. Following this approach, three different levels of data processing can be identified.

The first level is executed directly on a smart sensor that, although provided with limited resources, can manipulate the signals collected from the environment, digitalize them and "wrap" the raw data for further processing. In some cases, depending on the resources available on the sensor and the nature of the acquired signal (in terms of sample rate and dynamic field), it is also possible to identify simple events directly on the edge.

The second level of data processing is executed on more powerful devices, such as multiservice gateways, that are deployed near the sensor network and, following the edge computing paradigm, provide high-performance computing on the field. These devices are supplied with more hardware resources, with respect to the smart sensors, and can store and analyse wide temporal windows of the collected time-series. At this stage, the complexity of the algorithms increases significantly with respect to the complexity of the algorithms deployed at the sensor level. For many hardware platforms, the processing power allows to host embedded AI and to run containerized applications developed on the enterprise (see, for example Java OSGi technology or Microsoft Azure IoT Edge).

The third level is the "classical" big data processing, based on the high-performance computing capabilities available on cloud platforms or on data centers, at the enterprise level. Data processing is performed on the full datasets composed of all the collected time-series and can generate valuable insights and abstract information that feeds enterprise services (dashboards, production monitoring and control, predictive and preventive maintenance, processes yield, etc.).

### A. Sensor Network, Edge and Cloud Data Processing

The industrial use cases developed in CPS4EU (industry, automotive, etc.) adopt all the three levels of data processing previously described. Depending on the specific vertical application, the balance between them is achieved by partitioning the analytics functionalities in different algorithms that are tailored for the hardware on which they will run. This solution ensures efficiency in data processing, increases the control on the monitored processes, providing new low-latency functionalities and significantly reducing the bandwidth required for data transmission.

The opportunity to distribute data processing is given by the emerging paradigm of edge computing and the availability of sensors capable of executing low complexity algorithms.

in the cloud. This configuration mitigates the computational needs, splitting the processing between sensors and gaeteway, and reducing the network traffic load (see Figure 1, sensor 3).

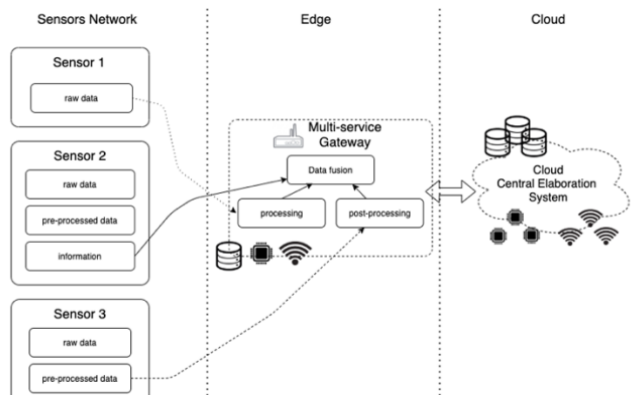

*Figure 1. Different configuration for a smart transducer network using edge and cloudcomputing.*

Depending on the specific application and hardware, platforms and methods, data processing distribution could be optimized by quantifying data losses and comparing the performance versus the communication tradeoff. However, to optimize an algorithm for the particular application or hardware is totally in contrast with the main goal of the family of standards for smart transducers, intended to achieve device and data interoperability. The standard ISO/IEC/IEEE 21451.001 must be adopted to obtain a system capable to effectively function and communicate. Conceived to develop standard software and hardware for smart transducers, this family of standards defines a set of common communication interfaces between sensors. It also establishes sensor network implementation regardless of the physical communication channel. The purpose is to enhance the computational power of smart transducers and facilitate the flow from sensor raw data processing to sensor information extraction and fusion by promoting pre-processing on smart transducers.

In the context of Industrial IoT and CPSs, the role of Smart Sensors equipped with pre-processing techniques is to contribute to part of the computational power that in other cases would be demanded to a remote central computing unit, reducing the load on the cloud system and on the communication network itself. To adopt this paradigm, smart transducer resource constraints need to be taken into account, such as: low computational capabilities, limited storage, short battery life and inadequate communication ability. Moreover, algorithms must be flexible, depending on the sensors and the specific application. So, the standard is structured in order to make different possible configurations:

1) The classic configuration where there is no data processing on sensors and all data is sent to the edge and, eventually, to the cloud, for processing. In this configuration, there are low computational load constraints on sensors, whereas there is a big traffic load of data in the network (seeFigure 1, sensor 1);

2) A configuration where sensors pre-processed data and results (i.e., information) are directly sent to the edge device and, eventually, to the cloud, for data fusion. In this configuration, there is an increased computational load for the sensors (which generally entails cost and complexity) but a decreased load for the network (see Figure 1, sensor 2);

3) A third configuration, where data is pre-processed by sensors and post-processed on the edge and, eventually,

For each described configuration, an edge device, acting as an intermediate level, is placed between the sensor network and cloud layers following the edge-computing paradigm (as described in the previous section). The objective is to improve processing capabilities without excessively loading the sensors or the central system, to distribute the computational loads in a more effective way and to decrease the network overload. By using the hardware needed by the IEEE21451 Transducer Interface Module (TIM), which requires a system based on a micro-controller, it is possible to implement data treatment services outside or directly on the TIM, providing extracted features rather than raw data for further analysis and processing. The purpose of this pre-processing technique is to extract basic information from the raw signal, the same information that a human being could provide looking at the signal itself. This technique emulates a human observer in the sense that it correlates samples to infer a global behavior, detecting, for example, a minimum or a maximum, or inferring signal shape parameters. In CPS4EU, Industrial Automation use cases will be equipped with a CPS compliant smart transducer and adopt the edge computing paradigm for addressing the different contexts and scopes implementing distributed sensing systems.

*B. Logical and Physical Architecture*

The proposed logical architecture develops a mixed configuration where smart sensors, edge computing units, and cloud services coexist, according to configuration option 3 (see paragraph A). Smart sensors equipped with algorithms that require low computational power, communicate through a secure and standard protocol with edge devices. When low latency is required, the gateway directly computes the information received from the field by the smart transducers and replies. Otherwise, the data is sent to a cloud service to be further elaborated and stored.

Regarding the Smart Transducer, we select and use the ESP8266 standard platform, a low-cost microcontroller unit (see Figure 2) which provides in a small package all the necessary features to implement both data acquisition and algorithm processing. The microchip is also equipped with a Wi-Fi communication interface to connect with edge devices (e.g. a gateway). In Figure 3, the logical implementation of this Distributed Smart Measurement System is shown. Several quantities need to be measured in the industrial environment, each with a different requirement in terms of sampling rate and resolution: e.g. temperature, pressure, humidity, distance. Moreover, when adding a new sensor to a piece of machinery, it is mandatory to consider its impact on the machine itself and on production operations. For this reason, the sensing element or transducer adopts shapes and connection interfaces that are application-dependent. In the case of an analog sensor, a conditioning circuit is mandatory to adapt the nature of its output and the voltage levels to the input range of the Analog to Digital Converter (ADC) of the microcontroller unit. If a digital unit is available to measure the required quantity from the environment or the machine, its output needs to be interpreted and converted into data for the gateway or the cloud service. The microcontroller executes a first conversion stage between the raw digital data converted internally from the ADC or digitally sent by a digital sensor. On these raw samples, simple algorithms can be applied to off-load the gateway and the communication channel they are connected with. The amount of computation made by the Distributed Smart Measurement System and the data sent to the gateway becomes a tradeoff that dictates the lifetime of the sensor.

Using a dedicated and secure Wi-Fi network, the pre-elaborated data is sent by the sensing node to the Edge Computing Gateway, according to the standard IEEE 21451. In this implementation, the smart transducer encapsulates the data in a JSON message standardized among all the sensors and the gateway. The standard message contains not only the data but also the time at which the sample has been collected, enabling the possibility of synchronization between distributed nodes [14-15]. In the simplest case, the sample consists of a single measure, converted by the microcontroller, or in a more sophisticated case it can be the result of the elaboration of a set of raw samples (for example the rotational frequency extracted from the analysis of the vibration of a drill [16]). This JSON message is then submitted to the MQTT broker running on the edge gateway on a specific topic. Each measurement has its destination, and the same measurement module can publish on different sub-topics. Compared with the the IEEE 21451 standard specification, the topic takes the role of the TIM because it incapsulates the information regarding the nature of the measure. When the edge unit or any other sensor requires a specific field, it can subscribe to the topic initialized by the sensors that measures that specific quantity. The intrinsic bidirectionality of this protocol allows the inter-communication between sensors and edge units.

V. EDGE COMPUTING GATEWAY

The proposed solution relies on a industry-grade edge computing platform, based on a high reliability industrial gateway and a software framework for edge computing and IoT (Eclipse Kura/Eurotech ESF).

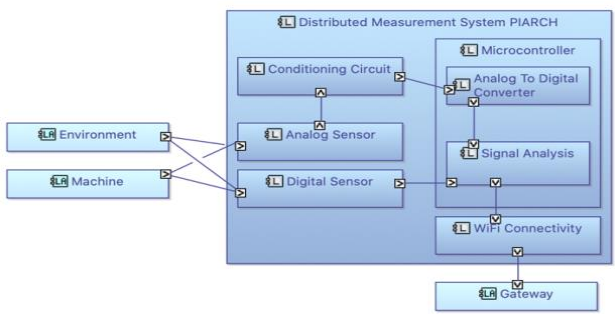

*Figure 2. Blocks diagram of a Smart Sensor Module.*

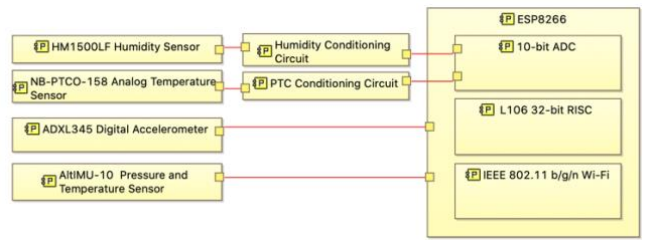

*Figure 3. Physical architecture of a typical Smart Sensor Module.*

The Edge Computing Gateway follows an IoT service gateway model, based on a multiservice framework running an enterprise-level software stacks and operating on the edge of an IoT deployment as an aggregator, orchestrator and controller. This model provides the best foundation for Edge Computing oriented solutions capable to shift data processing on the edge of the network, to reduce the communications bandwidth and to allow knowledge generation at or near the source of the data. The framework, based on Eclipse Kura (Eurotech ESF), is an Internet of Things (IoT) solution based on Java/OSGi and conceived to simplify the design, development, deployment, operation and maintenance of applications on the edge. It has been adopted as it easily allows to port applications on different hardware architectures, unifying the application engineering process, simplifying the cooperation between heterogeneous devices, and ensuring the dynamical evolution of the edge applications. Moreover, being an open source solution, it prevents vendor lock-in and guarantees the protection of the software investment.

Eclipse Kura runs on the edge on top of the Java Virtual Machine (JVM) hosted on a multiservice gateway by Eurotech. Kura is used as an IoT framework running on the gateway hardware to implement the data collection and processing logical components, which are implemented using Kura Wires. This component of Kura allows to easily implement the business logic on the edge by graphically designing the data collection flow and partitioning the processing between different modules, while interfaces are available to monitor the dataflow process. Moreover, Kura features a device hardware abstraction and a cloud platform abstraction that, on one side, simplify the interaction with the distributed measurement systems implementing the sensing nodes and, on the other, simplify the data analysis on the enterprise based on Azure Stack technology.

VI. CASE STUDY: TRIMMING QUALITY IMPROVEMENT

The case study deals with a CPS aiming to support the monitoring of a trimming production process on the aircraft composite part. During trimming/milling on composite materials of an aircraft, defects like delamination can occur, due to different phenomena that are difficult to manage because of the high complexity and high numbers of variables affecting the process (vibration, detachment of the part being cut, tool wear & speed, humidity, temperature, air pressure, etc.).

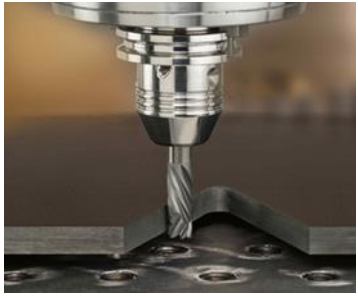

*Figure 4 Industrial trimming of a composite part.*

The objective of this use case is to introduce a CPS able to collect data coming from sensors and numerical control machines (CNC), to analyze the trimming process with a quality statistic algorithm and understand the main causes of defects, and eventually provide real-time information in order to change the setting of the trimming machine parameters to reduce the risk of damages or defects.

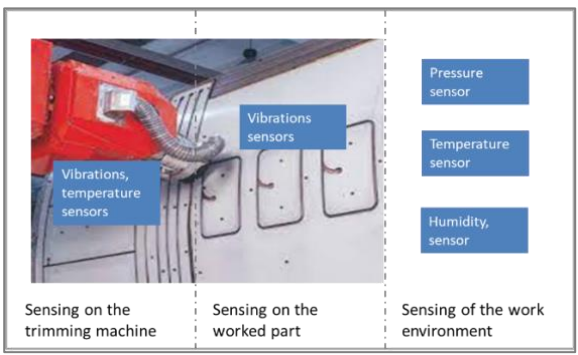

*Figure 5 Sensing in the trim & drill area (source Jobs)*

The use case deals with two phases:

1. The first phase focuses on the production of the quality prediction model. It starts from the acquisition of sensor data on trimming sessions, along with the corresponding ex-post evaluation of the output quality, in order to build a significant history to feed the AI/ML algorithms with the aim to identify the correlation between the parameters that can produce defects.

2. The second phase is devoted to the dynamic application of the quality prediction model on the edge, in order to suggest the setting of tool parameters, with the aim to reach the best final quality.

The architecture of the use case is partitioned in the following layers (see Figure 7):

- a perception layer where sensor signals are converted into a time series of data representing measurements of the relevant process variables, with appropriate sampling and enforcing data quality;
- an Edge Computing Gateway, hosting the Eclipse Kura IoT integration framework, is responsible of collecting the data streams from the distributed sensing nodes and sending them to the remote enterprise data analysis platform; it also runs on the edge the prediction model as a containerized Azure application; the model estimates the production performance and the risk of defects that is displayed on an HMI to the trimming operator who can decide to adjust the trimming machine settings;

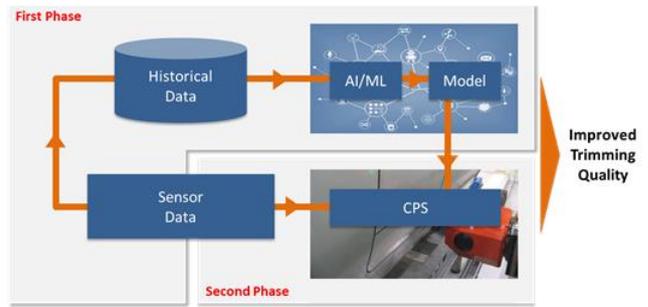

*Figure 6 Use case phases.*

- an enterprise data analysis platform (Leonardo Secure Connected Factory) that supports the data scientist in discovering correlations between trimming variables and defects found on the trimming output.

Several distributed smart measurement nodes are deployed on the trimming area to acquire various parameters, e.g. the vibrations of the trimming head and of the worked parts, the temperature and speed of the airflow from the trimming machine, the temperature and pressure of the working environment.

The smart measurement nodes in the sensing layer have been designed according to the reference distributed sensing architecture outlined in section IV. Depending on the parameter to measure, the node is equipped with a specific sensor and is conveniently powered. More in details, the following aspects are measured:

- Tri-axial vibration is collected for each Work Part (mounted on the composite part under the extra material that will be cut) and for the Support Mandrel. The sensing nodes are equipped with the vibration sensor analog devices ADXL 345. The power supply is provided by the LiPO battery, type: 18650.

- The air flux and temperature of the vacuum are measured in the trimming cap. The sensing node is equipped with the flux sensor Debimeter Bosch, which also provides the temperature.

- The sensing node deployed to measure the trimming head vibrations or the work part vibrations adopts a tri-axial accelerometer mechanically coupled with the machine or the carbon fiber part. The power supply is provided by the LiPO battery, type: 18650.

- Environmental parameters are collected by equipping the sensing node with DHT22, for humidity and temperature measures, and Pololu AltIMU-10 v4 for pressure. The power supply is provided by 8-24 V ac/dc.

Thanks to the onboard microcontroller on each smart measurement node, the devices acquire and process the signals. All the measures are pushed to an MQTT topic and collected during the trimming process by the Edge Computing Gateway. The envelope of each measurement station is made with PLA (Polylactic Acid) with 3D printers.

VII. CONCLUSIONS AND FUTURE WORKS

The analysis of the Industry Automation use cases included in CPS4EU reveals heterogeneous requirements in terms of data computing. This work proposes a general architecture and a solution for a distributed perception system. The proposed solution adopts standards and an open source IoT framework for simplifying the integration and the deployment of the measurement systems along the production line. These systems collect data during the manufacturing process, analyse the data to create a quality model and run the model on the edge to dynamically improve the quality of the production process.

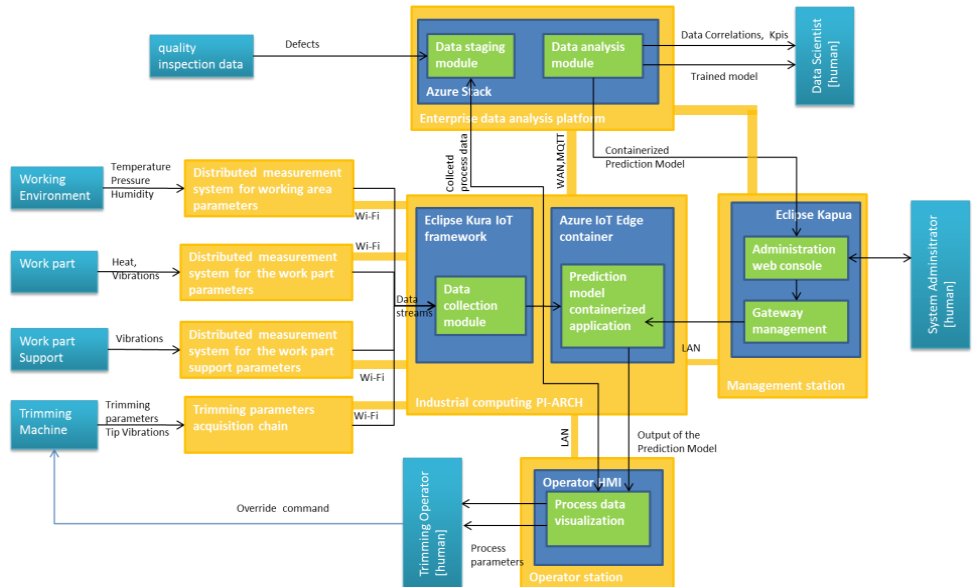

*Figure 7. Architecture of the use case.*

According to the mission of the project CPS4EU, the main contributions attain with the definition of pre-integrated architecture that is re-usable in different vertical applications and that has been and will be adopted in a real industrial use case, the Trimming Quality Improvement in the production process of a big part of a civil aircraft.

The next steps of the research activities will focus on the analysis of the historical data collected during the trimming process for the definition of the quality prediction model.

ACKNOWLEDGMENT

The CPS4EU project has received funding from the ECSEL Joint Undertaking (JU) under grant agreement No 826276. The JU receives support from the European Union's Horizon 2020 research and innovation programme and from France, Spain, Hungary, Italy, Germany."

The article reflects only the author's view. JU is not responsible for any use that may be made of the information it contains.